\documentclass[twocolumn]{aastex63}
\usepackage{graphicx}
\usepackage{subfigure}
\usepackage{multirow}
\usepackage{CJK}
\usepackage{mathptmx} 
\usepackage{hyperref}
\usepackage{lineno}
\hypersetup{linkcolor=blue,citecolor=blue,filecolor=cyan,urlcolor=blue}

\usepackage{textcomp}
\usepackage{paracol}

\begin{document}
        
\title{Network of Star Formation: Fragmentation  controlled by scale-dependent turbulent pressure and accretion onto the massive cores  revealed in the Cygnus-X GMC complex}

\author[0000-0003-3144-1952]{Guang-
Xing Li}
\affiliation{South-Western Institute for Astronomy Research, Yunnan University,
Kunming, Yunnan, 650500, P.R. China}

\author[0000-0002-6368-7570]{Yue Cao}
\affiliation{School of Astronomy and Space Science, Nanjing University, 163 Xianlin Avenue, Nanjing 210023, People's Republic of China}
\affiliation{Key Laboratory of Modern Astronomy and Astrophysics (Nanjing University), Ministry of Education, Nanjing 210023, People's Republic of China}
\affiliation{Center for Astrophysics $\vert$ Harvard \& Smithsonian, 60 Garden Street, MS 42, Cambridge, MA 02138, USA}

\author[0000-0002-5093-5088]{Keping Qiu}
\affiliation{School of Astronomy and Space Science, Nanjing University, 163 Xianlin Avenue, Nanjing 210023, China}
\affiliation{Key Laboratory of Modern Astronomy and Astrophysics (Nanjing University), Ministry of Education, Nanjing 210023, China}

\correspondingauthor{Guang-
Xing Li, Yue Cao, Keping Qiu}
\email{gxli@ynu.edu.cn} \email{caoyue@smail.nju.edu.cn} \email{kpqiu@nju.edu.cn}



\begin{abstract} 
  {Molecular clouds have complex density structures produced by processes including turbulence and gravity.}
  We propose a triangulation-based method to dissect  the density structure of a
molecular cloud and study the interactions between dense cores and their
environments. In our {approach}, a Delaunay
triangulation is constructed,  which consists of edges connecting these cores. Starting from this construction, we study the physical connections between neighboring 
 dense cores and the ambient environment in a systematic fashion. 
We apply our method to the Cygnus-X massive GMC complex and
 find that the core separation is related to the mean surface
density by $\Sigma_{\rm edge} \propto l_{\rm core }^{-0.28 }$, which can be
explained by {fragmentation controlled by a scale-dependent turbulent pressure (where the pressure is a function of scale, e.g. $p\sim l^{2/3}$)}.
 We also find that the masses of low-mass cores  ($M_{\rm core} < 10\, M_{\odot}$) are determined by fragmentation, whereas massive cores ($M_{\rm core}
> 10\,  M_{\odot}$) grow mostly through accretion. The transition from
fragmentation to accretion coincides with the transition from a log-normal 
core mass function (CMF) to a power-law CMF.  By constructing surface density profiles measured along edges that connect
neighboring cores, we find evidence that the massive cores have accreted a significant fraction of gas from their surroundings and thus depleted the gas reservoir. Our analysis reveals
a picture where cores form through fragmentation controlled by scale-dependent turbulent pressure
support, followed by accretion onto the massive cores, {and the method can be applied to different regions to achieve deeper understandings in the future.}
\end{abstract}


%
%

\section{Introduction}
Stars in the universe come from the collapse of cold and dense
molecular gas. The collapse is a complex process characterized by a multiscale
interplay between gravity, turbulence \citep[e.g.][]{2004RvMP...76..125M},
magnetic field \citep{2014prpl.conf..101L}, ionizing radiation
\citep{2011MNRAS.414..321D}, as well as the Galactic shear
\citep{2014prpl.conf....3D,2016A&A...591A...5L}. Despite years of efforts, a clear picture of the
 importance of these ingredients has not been agreed upon.

In star formation study, gas structures of  sizes of around 10 pc are called
clouds. Within a cloud, pc-sized structures are called clumps and 0.1
parsec-sized structures are called dense cores. It is
believed that the collapse of clumps leads to the formation of star clusters
\citep[e.g.][]{Williams2000}, and the collapse of  dense cores leads to the
formation of either  individual stars \citep[e.g.][]{2007A&A...462L..17A} or binaries/multiple systems \citep{2005A&A...439..565G}.

Dense cores are the direct precursors to stars.
Understanding how dense cores acquire mass is thus crucial for
understanding the origin of the stellar mass. In particular, the formation of cores with masses
of a few tens of solar masses are of particular interest since they are likely
to collapse to form massive stars -- the type of star that is rare in number contributes to a large portion of the light we observe in the universe
\citep[e.g.][]{2007A&A...476.1243M,2007prpl.conf..165B,2017A&A...602A..77T}
. Dense cores are formed as the result of cloud collapse. After formation, the
 dense cores will exchange energy and momentum with their environments, i.e.
 accretion should lead to growth in the core mass, where turbulence can be driven simultaneously 
 \citep{2012ApJ...750L..31R,2015ApJ...804...44M,2018MNRAS.477.4951L,2020arXiv200904676G}. 
 It is thus necessary to study the interaction between dense cores and 
 their environments systematically.  We propose a novel 
 method to achieve this, where analyses are achieved using the Delaunay triangulation algorithm
 \citep{delaunay1934}.

\section{Cygnus-X GMC complex}
\begin{figure*}[htb!]
\epsscale{1.1}\plotone{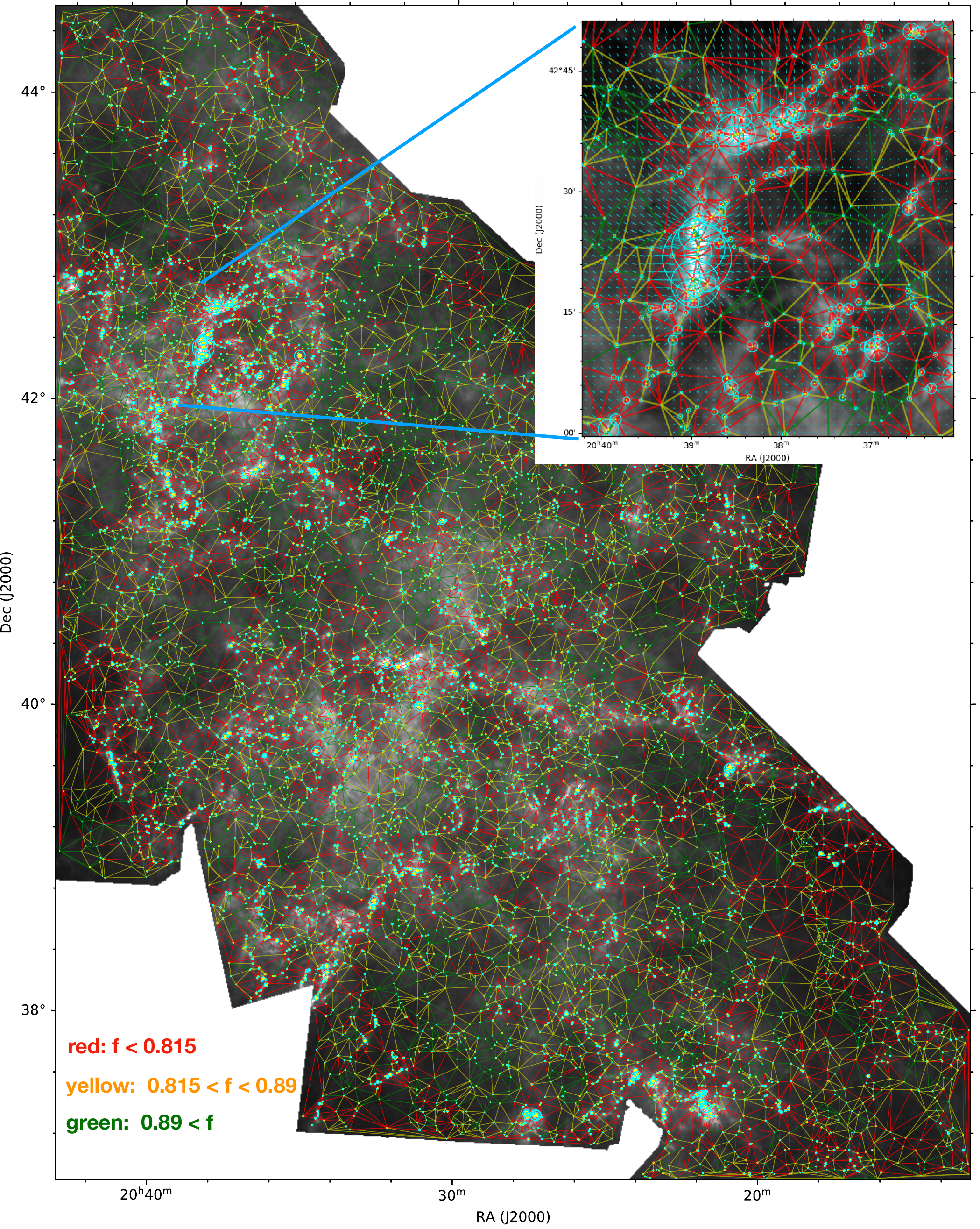}
 \caption{\label{fig:mesh} {Delaunay triangulation} constructed on the Cygnus-X GMC complex. The background image is a surface density map
 of the region constructed  using  data from  \citet{2010A&A...518L..77M}. On top of it, we overplot the dense core sample presented in Cao et al. (2021 {under review}), as well as the corresponding Delaunay
 triangulation diagram. {The edges of the triangulation diagrams are colored according to their filling factors} (Eq.(\ref{eq:filling})): the red lines stand for edges with filling factors smaller than 0.81, the yellow
 lines stand for edges with filling factors $0.815 < f < 0.89$, and the green  lines stands for filling factors larger than 0.89. See Sec.
\ref{sec:filling} for details. {The accretion radii are estimated using Eqs. (\ref{eq:acc:1}) and (\ref{eq:acc:2}) and are indicated using circles.}
In the inset plot, we present a zoom into the DR21 massive star-forming region, where we  overplotted the acceleration vector estimated using the method
 presented in \citet{2016arXiv160305720L}.  }
\end{figure*}

The region we study is the Cygnus-X Giant Molecular Cloud (GMC) complex. Located at a distance of 1.4 kpc \citep{2012A&A...539A..79R},  it is one of the most well-studied active star-forming regions. 
The surface density
structure of the region has been constructed by \citet{2019ApJS..241....1C}, as part of the CENSUS project, using data from the HOBYS survey \citep{2010A&A...518L..77M} obtained with the Herschel telescope. The surface density map has a resolution of 18$^{\prime \prime}$.4 ($\approx$ 0.13 pc).

Based on the surface density map, a dense core sample has been constructed by
Cao et al. (2021 {under review}) where they extracted sources from the surface
density map using the \texttt{getsources} software \citep{2012A&A...542A..81M}.
Source properties such as coordinates, full-width-at-half-maximum (FWHM) diameters, peak column densities, and masses are derived by fitting two-dimensional Gaussian models to the data.
The final version of the sample contains  8431 objects, where additional criteria such as source size and mass uncertainty 
are applied to ensure robust detections. The dense cores in the sample have
masses that range from 0.1 to 10$^3$ $M_{\odot}$.  

In Fig.  \ref{fig:mesh}, we plot the surface density distribution of the region, where substructures of vastly different scales can be identified. Filamentary structures
prevail \citep{2014prpl.conf...27A} in the map. Dense cores from Cao et al. (2021 under review) are overlaid.

 Gravity is the  driver
 of star formation. To visualize its effect, following
 \citet{2016arXiv160305720L}, we constructed the gravitational field of the
 region and overplotted the acceleration vectors on the map. 
 The potential is computed assuming that the mass distributes on a slab with a thickness of $\sim 0.3\;\rm pc$.
 The  gravitational acceleration  exhibits a complex pattern, and gravity from the very massive cores seems to be able to exert influence on a
 scale ($\gtrsim 1$ pc) that extends much beyond its physical size. 

%
%
%


%

\section{Method}\label{sec:method}
\subsection{Constructing the triangulation mesh}
To address the connection between dense cores and the environment, we construct a Delaunay triangulation mesh that
consists of a set of edges connecting a set of points, upon which detailed analyses are formed. The detailed procedure is outlined below:
\begin{enumerate}
  \item Obtain the data. Our data consist of a map of the surface density structure of the Cygnus-X region, and a catalogue of dense cores. 
  \item Construct the Delaunay triangulation mesh. We use the \texttt{Delaunay} function from the \texttt{Scipy} package \citep{2020SciPy-NMeth} to construct the mesh. The input consists of the locations of dense cores, and the output is a mesh, which consists of edges that connect the dense cores. 
  \item Derive the properties of the edges. An edge that connects one dense core $p_1$ to another core $p_2$ should contain the following attributes: mass of the dense core at $p_1$, $m_1$, and  mass of the dense core at $p_2$, $m_2$, the length {(the separation between adjacent cores)} $l_{\rm core}$, the mean surface density of the edge, defined as $\Sigma = \int_{p_1}^{p_2} \Sigma(x) {\rm d} x / l_{\rm core} $, and the filling factor $f$ defined in Eq. (\ref{eq:filling}). 
  \item Study the correlations between attributes. 
\end{enumerate}

\begin{figure}[htb!]
  \epsscale{1}\plotone{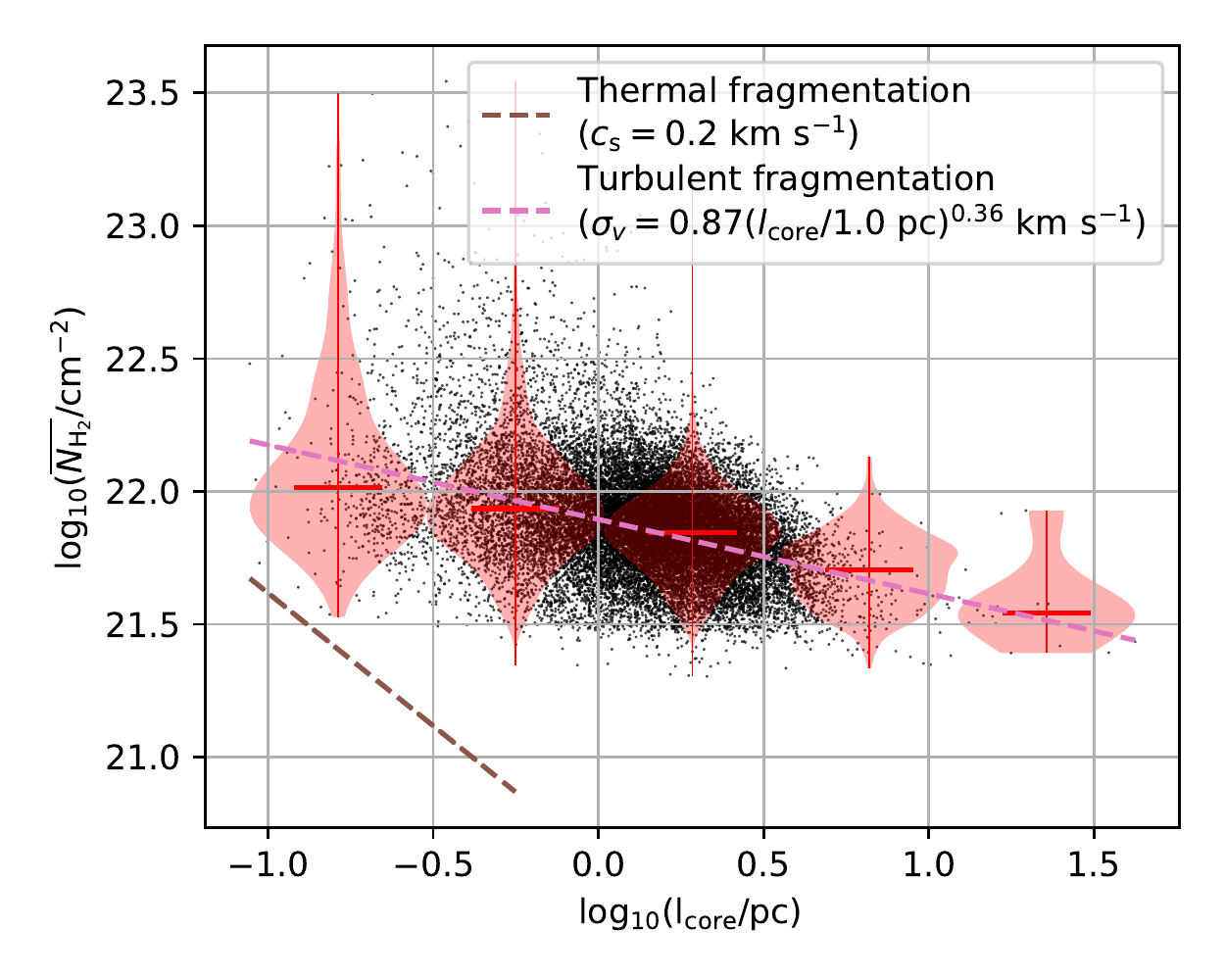}

  \caption{\label{fig:turbulence}  {The role of turbulent pressure support in fragmentation}. We plot the relation between projected core separation $l_{\rm core}$ and surface density $\Sigma_{\rm gas}$, where $l_{\rm core}$ is measured from the length of the edges, and 
  $\Sigma_{\rm gas}$ is the mean surface density  measured on these edges.  The violin plot (in red) outlines illustrate kernel probability density, i.e. the width of the shaded area represents the proportion of the data located there. Different lines represent predictions from different fragmentation scenarios, where the brown dashed line stands for the case where thermal support is effective (or the cases where a scale-independent support is effective), and the pink dashed line represents the case for a scale-dependent turbulent support. See Sec. \ref{sec:turbulence} for details  
  }
\end{figure}
  
\begin{figure*}[htb!]
  \epsscale{0.53}\plotone{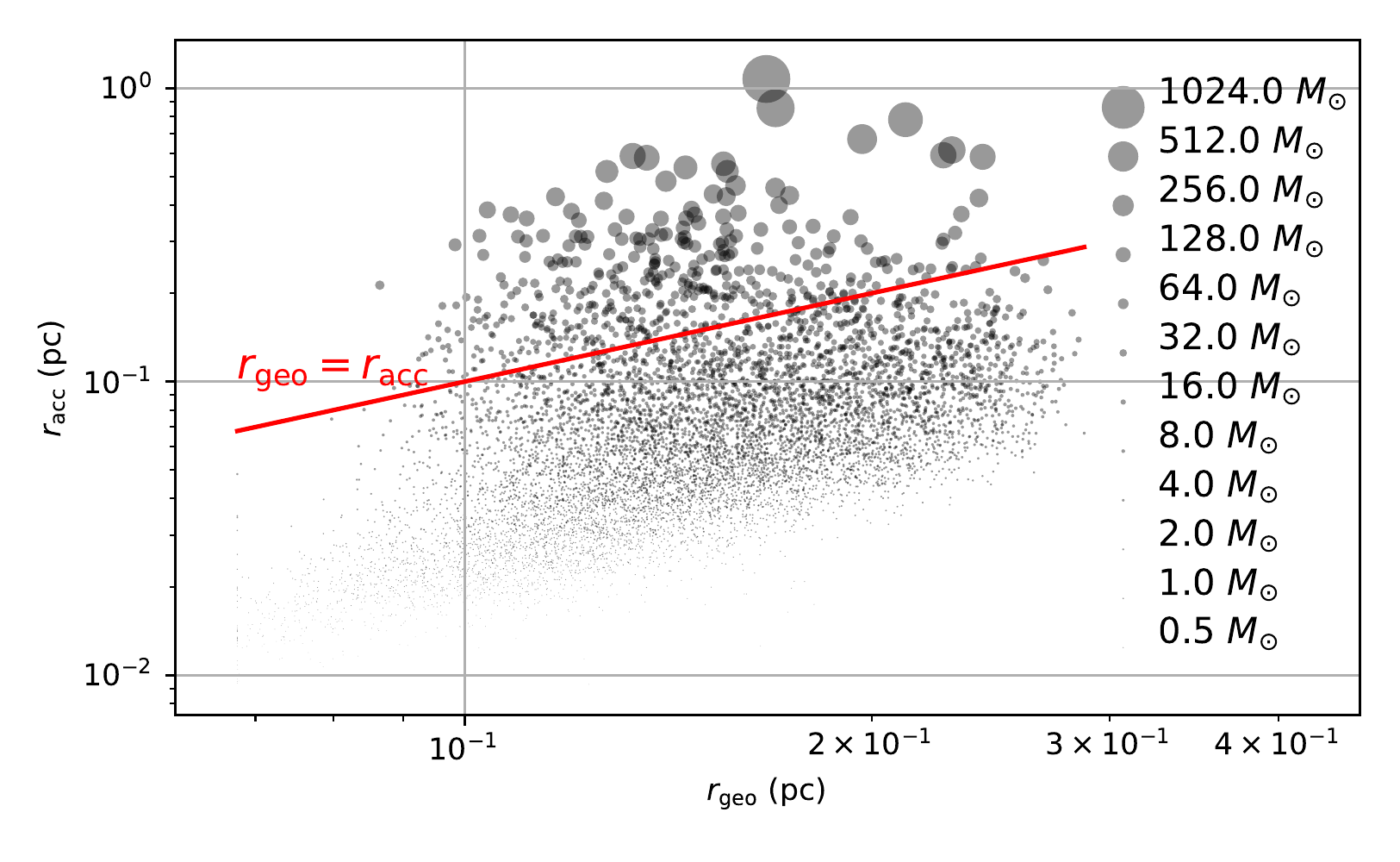}
  \epsscale{0.41}\plotone{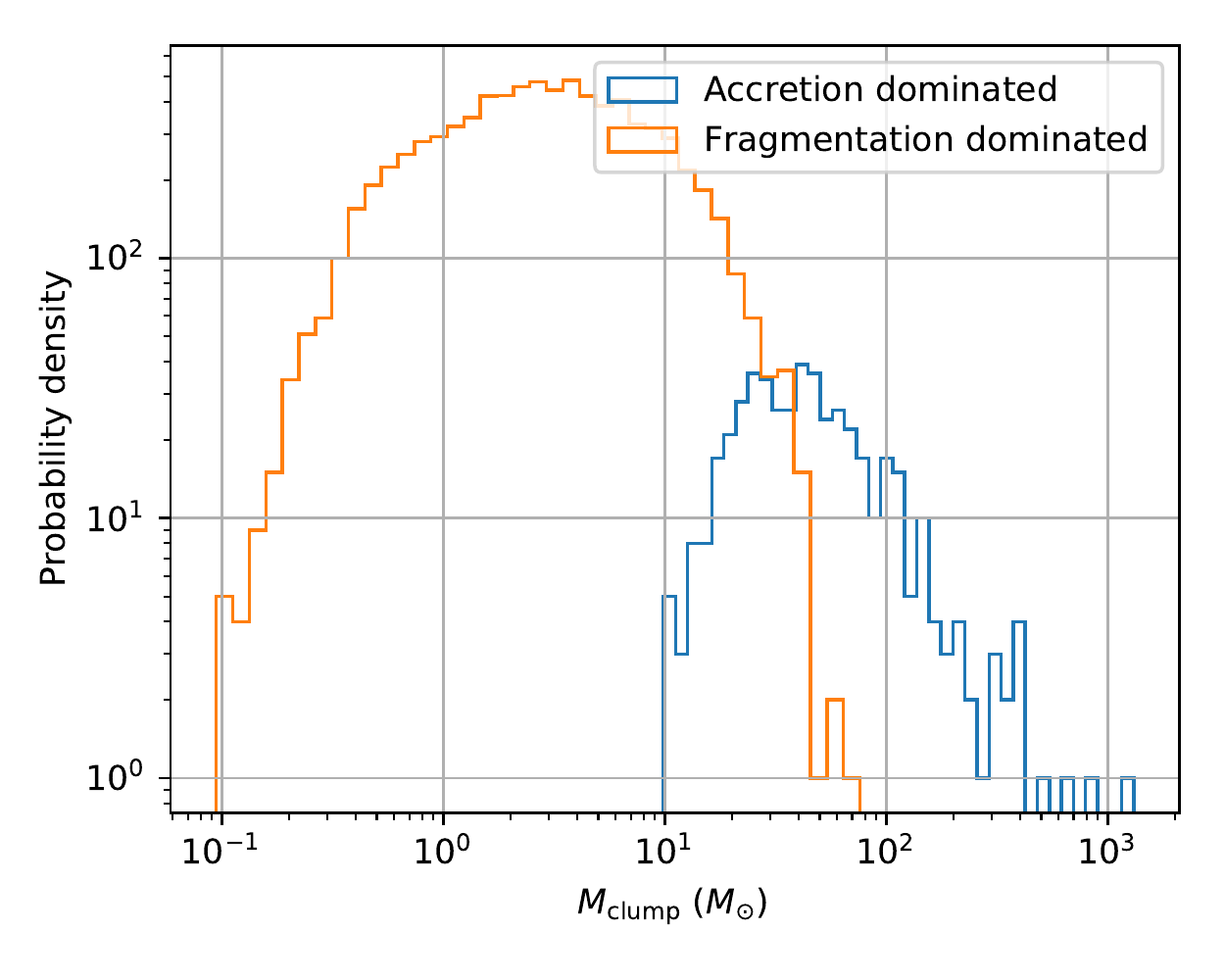}
 
  \caption{\label{fig:CMF}  {Effect of accretion on core growth}.
{Left:} Distribution of cores in the $r_{\rm goe}$-$r_{\rm acc }$ plane. $r_{\rm geo}$ is the core size measured by Cao et al. 2021, and $r_{\rm acc}$ is the sum of the accretion radii of these two cores. Disks of different sizes represent cores of different masses.  The red line marks the region where $r_{\rm geo} = r_{\rm acc }$. Cores with  $r_{\rm geo} > r_{\rm acc }$ are called ``fragmentation dominated", whereas cores with  $r_{\rm geo} < r_{\rm acc }$ are ``accretion dominated". 
  {Right:}  Mass Function of different Core Groups, we plot the CMF (Core Mass Function) for cores that belong to Accretion-dominated regime, the fragmentation dominated regime. Low-mass cores ($M_{\rm core} < 10\; M_{\odot}$) are fragmentation dominated whereas high-mass cores ($M_{\rm core} > 10\; M_{\odot}$) are accretion dominated.  }
 \end{figure*}

 \section{Results \& Discussions}
 \subsection{Separation versus Surface Density : Evidence for  scale-dependent turbulent fragmentation} \label{sec:turbulence}

Dense cores are the very
product of cloud fragmentation. The spatial distribution of the
dense cores contain information on how the cloud fragments. Based on our triangulation mesh, we plot the separations between
dense core against the mean surface densities, where the projected separations $l_{\rm core}$ are measured from the length of the edges, and 
$\Sigma_{\rm gas}$ are the mean surface densities measured along these edges in Fig. \ref{fig:turbulence}. A fit to the data suggests the following relation:\footnote{We performed our fit in the log-log space, and find ${\rm log_{10}}(N_{\rm H_2}/ {\rm cm^{-2}}) =  (-0.29 \pm  0.004)\times {\rm log}_{\rm} (l_{\rm core} / 1 {\rm pc} )  +  (21.9\pm 0.0013) $.}
\begin{equation}\label{eq:fitting}
N_{\rm H_2} \approx 8.1 \times 10^{21} {\rm cm^{}}\; (l_{\rm core} / 1\, {\rm pc})^{-0.28}\;.
\end{equation}

This relation demands an explanation. To proceed, we note that due to
observational constraints, we are only able to measure the projected separation
between core pairs from the lengths of the edges. Using results from numerical
simulations, we have verified that the 2D projected separation can be used as a
proxy for the 3D separation to good accuracy. The details can be found in
Appendix \ref{sec:appen:comp}.

We aim to derive the relation between core separation and surface density in a
scenario that involves the interplay between pressure support and gravity. The
support can take the form of thermal pressure as well as turbulent pressure.
Thermal pressure is effective in supporting against the collapse of small
clouds, such as isolated globules \citep{2001Natur.409..159A}. Turbulent
pressure can be effective when the nonthermal velocity dispersion $\sigma_{\rm
turb}$ exceeds that of the thermal one. However, there are still controversies
concerning how turbulence pressure support takes effect. In models such as
\citet[][]{2003ApJ...585..850M}, the turbulence pressure support can be approximated
as a pressure $p_{\rm turb} \approx \rho \sigma_{\rm turb}^2$ where $\sigma_{\rm
turb}$ is a constant for a region, whereas  \citet[][]{Li2017,Zhang2017b}
pointed out that to explain the observed mass-size relation of star
cluster-hosting clumps, the turbulence pressure support must be {\it scale-dependent}
e.g. $\sigma_{\rm turb}\approx l^{1/3}$ where $l$ is the scale, and $p_{\rm
turb} \approx \rho {\sigma_{\rm turb}}^2 \approx l^{2/3}$.  

To explain the observed correlation found in Fig. \ref{fig:turbulence}, we consider two scenarios: First, we consider the case where the support is provided either by a thermal pressure  or by a scale-independent
turbulent pressure, and derived the expected relation between core separation and surface density for both cases. Using dimensional arguments, we find that in the cases where thermal support is effective, the core separation is connected to the surface density by 
\begin{equation}
  l_{\rm core, thermal} \approx \frac{c_{\rm s}^2}{G \Sigma_{\rm gas}}\,,
 \end{equation}
 and 
 \begin{equation}
\Sigma_{\rm gas}\approx \frac{c_{\rm s}^2}{G   l_{\rm core, thermal}}\,.
 \end{equation}
 In the case where a scale-dependent turbulence pressure support ($\sigma_{\rm v, turb } = \sigma_0 (l / l_0)^{ \beta}, \beta \approx 1/3)$ is effective, we have
 \begin{equation}
 l_{\rm core, turbulent} \approx \epsilon^{\frac{-2 \beta}{2 \beta - 1}} G^{\frac{1}{2 \beta - 1}} \Sigma_{\rm gas} ^{\frac{1}{2 \beta -1}}
 \;,
 \end{equation}
and 
\begin{equation}
\Sigma_{\rm gas} \approx \epsilon^{ 2 \beta} G^{ - 1} l_{\rm core, turbulent}^{2 \beta - 1}\;, 
  \end{equation}
where $\epsilon = \sigma_0^{1/\beta}  l_0^{-1}$.
 Details concerning the derivation of these equations can be found in Appendix \ref{sec:appen:ana}. 
 In the case of thermal support, we assume a sound speed of 
$c_{\rm s} = 0.2\;\rm km\, s^{-1}$ and in the case of scale-dependent turbulent pressure support, we assume $(\sigma_{\rm v, turb} / {\sigma_0}) = (l /l_0)^{\beta}$. To explain Eq. (\ref{eq:fitting}), we require  $\sigma_0 = 0.87 \rm \; km\, s^{-1}$, $\beta=0.36$. Note that our scaling relations are derived using dimensional arguments, and the value of $\sigma_0$ is not well-constrained. Nevertheless, both the values of $\sigma_0$ and $\beta$ are consistent with results from previous papers \citep{1981MNRAS.194..809L,2011ApJ...740..120R}. 

Thus, it is the scale-dependent turbulent support that determines how the cloud fragments into dense cores. {We note that under this framework, the turbulent pressure is a function of scale, e.g $p_{\rm turb} \approx \rho \sigma_{\rm v, turb}^2 \propto l^{2 \beta}$.}
On smaller scales, as the thermal pressure gradually dominates over the turbulent pressure, it is possible that the fragmentation is thermal in that regime, as indicated by some observations  \citep{2015MNRAS.453.3785P,2018ApJ...855...24P}, although turbulence can also be important \citep{2009ApJ...696..268Z,2018ApJ...855....9L}.

\subsection{Accretion as the origin of power-law CMF}\label{sec:cmf}
The origin of core mass remains an open question. Some theories predict that
cores acquire their masses  through turbulent fragmentation
\citep{2002ApJ...576..870P}. This  process can be described analytically, and
the resulting core mass function is log-normal-like
\citep{2009ApJ...702.1428H,2012MNRAS.423.2037H}. 

After formation, the  cores should be able to grow further by accretion gas from their surroundings.  The significance of this accretion remains to be acknowledged.  
To demonstrate this effect, we estimate the radius within which 
gravity a dense core is expected to be effective. Observations \citep[e.g.][]{1981MNRAS.194..809L} indicate that the velocity dispersion of molecular gas is a function of the scale, where approximately,
\begin{equation}
 \sigma_{\rm v}^{\rm turb} = \sqrt{\sigma_0^2  (l / l_0)^{2 \beta} + c_{\rm s}^2}\;,
\end{equation}
where $\beta \approx 0.38$. $\sigma_0 \approx 1 \;\rm km/s $, $l_0\approx 1\;\rm pc$ are normalizing factors  and $c_{\rm s} $ is the sound speed. The velocity dispersion caused by self-gravity can be estimated as
\begin{equation}
  \sigma_{\rm v}^{\rm turb} = \sqrt{G m_{\rm core}/ l}\;,
\end{equation}
and the accretion radius $l_{\rm acc}$ can be derived by solving  $ \sigma_{\rm v}^{\rm turb} = \sigma_{\rm v}^{\rm grav}$ \citep[e.g.][]{2012ApJ...746...75M} where
\begin{equation}\label{eq:acc:1}
  l_{\rm acc} = G m_{\rm core} / c_{\rm s}^2 {\; \rm when \; } c_{\rm s} < \sigma_{\rm v, turb}\;, 
\end{equation}
and
\begin{equation}\label{eq:acc:2}
  l_{\rm acc} = (G^{1/2} m_{\rm core}^{1/2} l_0^{\beta} \sigma_0^{-1})^{2/(2 \beta + 1)}{\; \rm when \;} c_{\rm s} <  \sigma_{\rm v, turb}\;.
\end{equation}
  $l_{\rm acc}$ is the radius within which the movement of gas would be affected by the presence of the dense cores.

{The accretion radius is dependent on the parameters in the Larson relations. Observations indicate that the slope of the Larson relation varies between 1/3 and 1/2 \citep{1981MNRAS.194..809L,2004ApJ...615L..45H,2011ApJ...740..120R}, with $\beta \approx 1/2$ been reproduced in some numerical simulations \citep[e.g.][]{2019A&A...630A..97C}. Observational studies also indicate that surface density can also be playing a role \citep{2009ApJ...699.1092H,2011MNRAS.411...65B,2019MNRAS.490.2648B}. 
Furthermore, \citet{2018ApJ...864..116Q} find that in clouds such as the Taurus molecular cloud, the relation  between core separation and velocity difference is better described by multiple power-laws, implying that the slope of the Larson relation might not be unique.  To compute the accretion radii, we make use of the ``optimal'' Larson relation, where $\sigma_0 = 0.87 \rm \; km\, s^{-1}$, $\beta=0.36$, as this is the version needed to explain the correlation found in Fig. \ref{fig:turbulence}. We also experimented with $\beta = 0.5$, and find that the differences are minimal.}
{The accretion radii are plotted in}  Fig. \ref{fig:mesh}. It can be readily seen that the 
expected sphere of influence of the most massive cores can extend well beyond their close vicinities such that gas accretion is expected to occur.

To evaluate under what conditions do accretion becomes important, in Fig. \ref{fig:CMF} we plot the accretion radii of the cores against the core sizes $l_{\rm geo} $ derived in Cao et al. (2021 {under review}) where the sizes are derived by fitting 2D-Gaussians to the data. From this plot, we draw the line of $l_{\rm acc} = l_{\rm geo}$, {and thus cores are thus divided into these two groups:
\begin{enumerate}
  \item fragmentation dominated, where 
    \begin{equation}\label{eq:l:frag}
      l_{\rm
      acc} < l_{\rm geo}\;,
    \end{equation}
    and
  \item accretion dominated, where,
  \begin{equation}\label{eq:l:acc}
    l_{\rm
    acc} > l_{\rm geo}\;.    
  \end{equation}
\end{enumerate}
Fragmentation-dominated cores have larger sizes, and are still in the stage of collapsing, whereas accretion-dominated cores are massive and centrally condensed, thus they should be accreting gas from their surroundings. } The CMFs of these core
groups are plotted in Fig. \ref{fig:CMF}. We find that the transition from the
fragmentation-dominated regime to the accretion-dominated  regime occurs at around $20\;
M_{\odot}$. This corresponds well to the transition from a log-normal CMF to the
power-law CMF found in Cao et al. (2021 under review) for the same region. Therefore, we propose 
$M_{\rm core} \approx 20\; M_{\odot}$ marks the boundary between
fragmentation-dominated growth and accretion-dominated growth for cores in the
Cygnus-X region.\footnote{We note that the results are dependent on the velocity-size relation. We considered (a) $\sigma_0 = 0.87 \rm \; km\, s^{-1}$, $\beta = 0.36$ and (b) $\sigma_0 = 0.87 \rm \; km\, s^{-1}$,  $\beta = 0.5$, and conclusions are largely unchanged. } The transition from fragmentation to accretion agrees with results from previous theoretical studies. It is believed that turbulent fragmentation creates a log-normal-like CMF \citep{2009ApJ...702.1428H,2012MNRAS.423.2037H}, after which the cores 
should accrete mass further which leads to formation of the power-law part of the CMF \citep[e.g.][]{2006MNRAS.370..488B,2012ApJ...746...75M}, {and this transition might eventually be related to the emergence of the power-law part of the density PDF found by previous studies \citep[e.g.][and references therein]{2000ApJ...535..869K,2009A&A...508L..35K,2011ApJ...727L..20K,2017ApJ...834L...1B}.  }

\subsection{The depletion of gas by massive cores}\label{sec:filling}
\begin{figure*}[htb!]
  \epsscale{0.4}\plotone{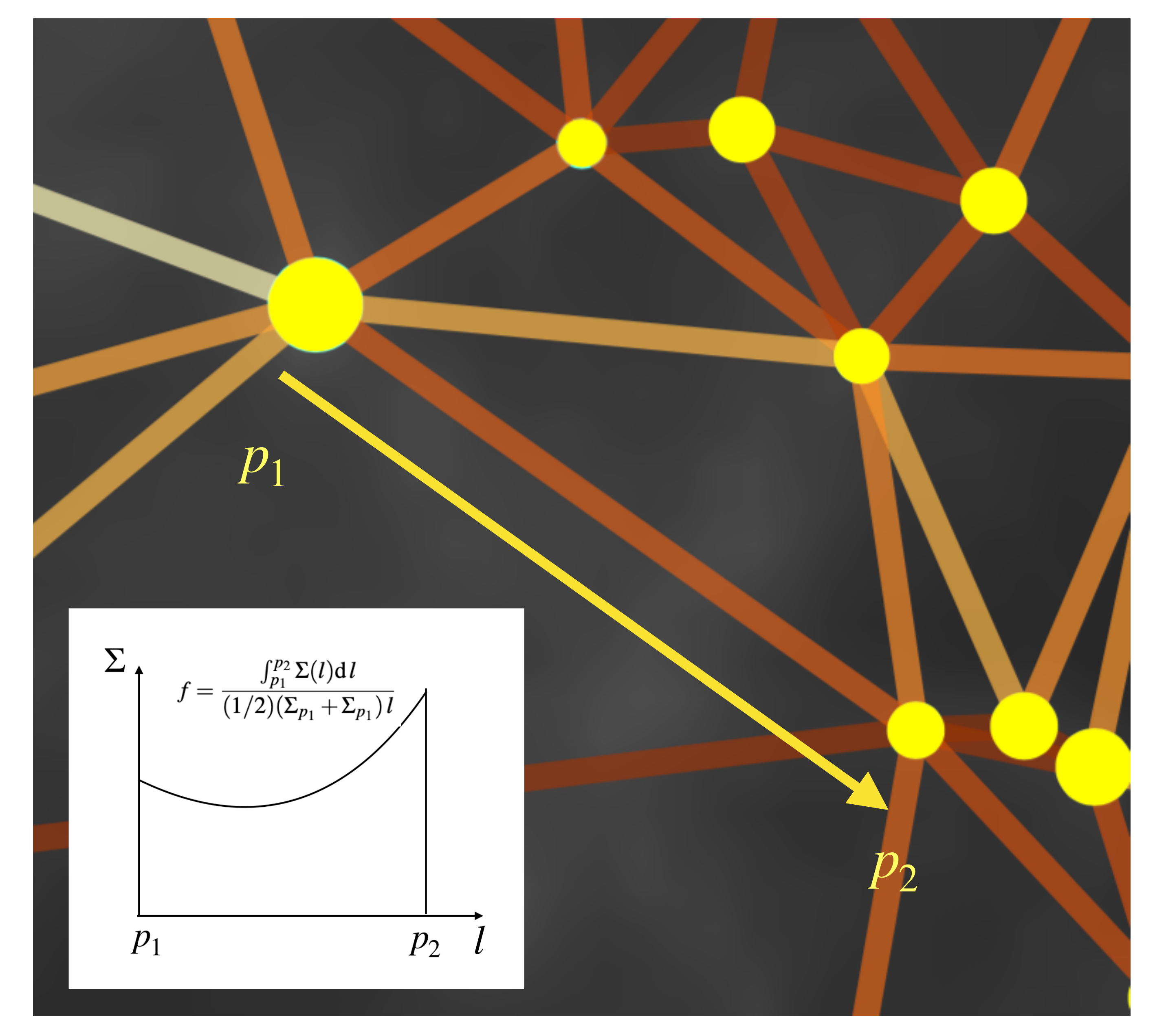}
  \epsscale{0.48}\plotone{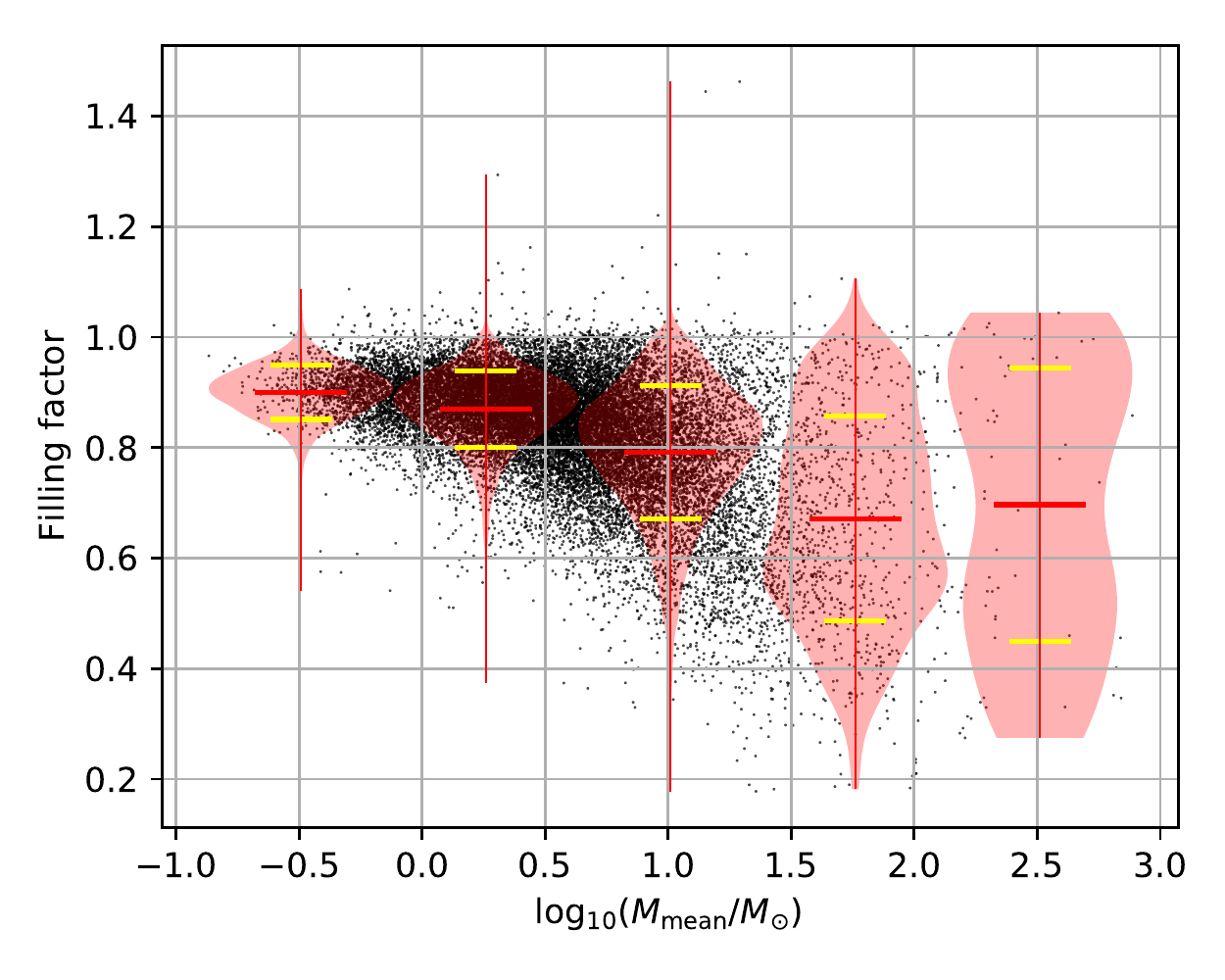}
  \caption{ {Relation between filling factor  Eq. (\ref{eq:filling} and core mass. }
{Left:} An illustration of the definition of the filling factor. The background image represents the surface density distribution of a sub-region. Yellow dots represent dense cores,  which are connected by edges.  To compute the filling factor of an edge that connects  $p_1$ to  $p_2$, we first construct a surface density profile measured along the edge, and then measure the surface densities at location $p_1$ and $p_2$. The filling factor is computed according to {Eq. (\ref{eq:filling})}. It is a measurement of the amount of gas found along the edge compared to the amount of gas one would expect if the region  is filled evenly.  
{Right:}
  Filling factor plotted against mean core mass. The violin plot (in red) outlines illustrate kernel probability density, i.e. the width of the shaded area represents the proportion of the data located there. The means and standard deviations of the distributions are indicated using the red and yellow horizontal bars, respectively.  The filling factor decreases as the core mass increases -- massive cores are connected by edges with low filling factors. \label{fig:filling} }
 \end{figure*}

Accretion of gas by the massive cores should lead to a depletion of gas from their surroundings. To quantify the depletion, we propose a ``filling factor", which, in essence, quantifies the concaveness of surface density profiles measured along edges that connect one core $p_1$ to another core $p_2$. To be more precise, the filling factor is defined as 
\begin{equation}\label{eq:filling}
 f  = \frac{\int^{p_2}_{p_1} \Sigma(l) {\rm d}\, l }{(1/2)
  (\Sigma_{p_1} + \Sigma_{p_1})\, l}\;, \end{equation} 
and this is illustrated in Fig. \ref{fig:filling}. {In our calculations, one edge consists of several pixels. The mean surface density of the edge is thus the mean of the surface densities measured at these pixel locations. We did not perform geometrical corrections to the line length, as it gets canceled out.}
According to this definition, if the gas located along the edge that connects two dense cores is evenly distributed, the filling factor should approach unity;
if the gas in the intercore region has been accreted by the dense cores, the density profile should have a concave shape, and the filling factor should be smaller than 1.  In Fig. \ref{fig:mesh}, we colored the edges according to their filling factors, where these ``depleted edges" -- edges with low filling factors -- can be identified in regions surrounding the very massive cores such as those on the DR21 filament.

Fig. \ref{fig:filling} plots the filling factor as a function of the mean mass $m_{\rm mean} = (1/2) (m_1 + m_2)$, where an increase
in core mass is accompanied by a decrease of the filling factor as well as an increase of
the dispersion of the filling factor. This can be taken as evidence that 
massive cores  (cores whose masses are larger than 10 $M_{\odot}$) have already accreted a significant amount of mass, which leads to the  depletion of their gas reservoirs, as seen from {the} decreasing filling factor.

\section{Conclusion}
We {develop} a new, triangulation-based method to study the relation between dense cores and their ambient environments in a systematic fashion. In our {method,} the locations of dense cores are taken as inputs, upon which  a Delaunay triangulation 
is constructed. {Based on the triangulation diagram, we study the relations between properties of cores such as core mass, core separation, and the density structure of ambient gas. }

We find that the core separation $l_{\rm core}$ is related to the surface density $\Sigma$ by $\Sigma \propto l_{\rm core}^{-0.28}$, which is inconsistent with the prediction of thermal fragmentation, but  can be well explained by fragmentation controlled by scale-dependent turbulent pressure  ({(e.g. the turbulent pressure is a function of scale, e.g. $p\sim l^{2/3}$), similar to the case of \citet{Li2017}.} 

{By proposing a criteria to distinguish between accretion and fragmentation (Eqs. (\ref{eq:l:frag}) and (\ref{eq:l:acc}))  , we find} that the masses of low-mass cores are likely to be determined by fragmentation. In {contrast} to this, accreting gas from the vicinities serves as a major way through which massive cores ($M_{\rm core} > 20\;M_{\odot}$)  acquire mass.  The transition between
fragmentation and accretion coincides with the transition from a log-normal 
core mass function (CMF) to a power-law CMF. By defining a quantity called the ``filling factor'' for the edges, we find  evidence that accretion onto the massive cores leads to a depletion of gas from their surroundings.

The density structures of molecular clouds are known to be extremely complex.  The method we developed captures this complexity, which allows us to further address the connections between cloud evolution, core {formation, and} growth in a systematic fashion.  {It is also possible to incorporate other information, such as radial velocity measurements, into our analyses.} By applying it to different regions and observations on different scales, deeper understandings {of the star formation process can}  be expected.  \\

\emph{Acknowledgments}
We thank our referee for a very careful reading of our paper and for the constructive
Guang-Xing Li acknowledges support from NSFC grant W820301904.
Yue Cao and Keping Qiu acknowledge supports from National Key R\&D Program of China No. 2017YFA0402600
National Science Foundation of China Nos. U1731237, 11629302, 11473011.
Y.C. is partially supported by the Scholarship No. 201906190105 of the China Scholarship Council and the Predoctoral Program of the Smithsonian Astrophysical Observatory (SAO). 
K.Q. acknowledges the science research grants from the China Manned Space Project with No. CMS-CSST-2021-B06.

\appendix
\section{Relation between 3D distance and 2D projected distance}
\label{sec:appen:comp}
Due to observational limitations, we are only able to map of the distribution of gas on the sky plane. As a result, only the core separation on the sky plane can be measured. In our theoretical analysis, we only concerned with the 3D separation. For our conclusions to be valid, we must address the connection between the 2D projected core separation and the true 3D core separation.

\begin{figure*}[htb!]
  \includegraphics[width = 0.5 \textwidth]{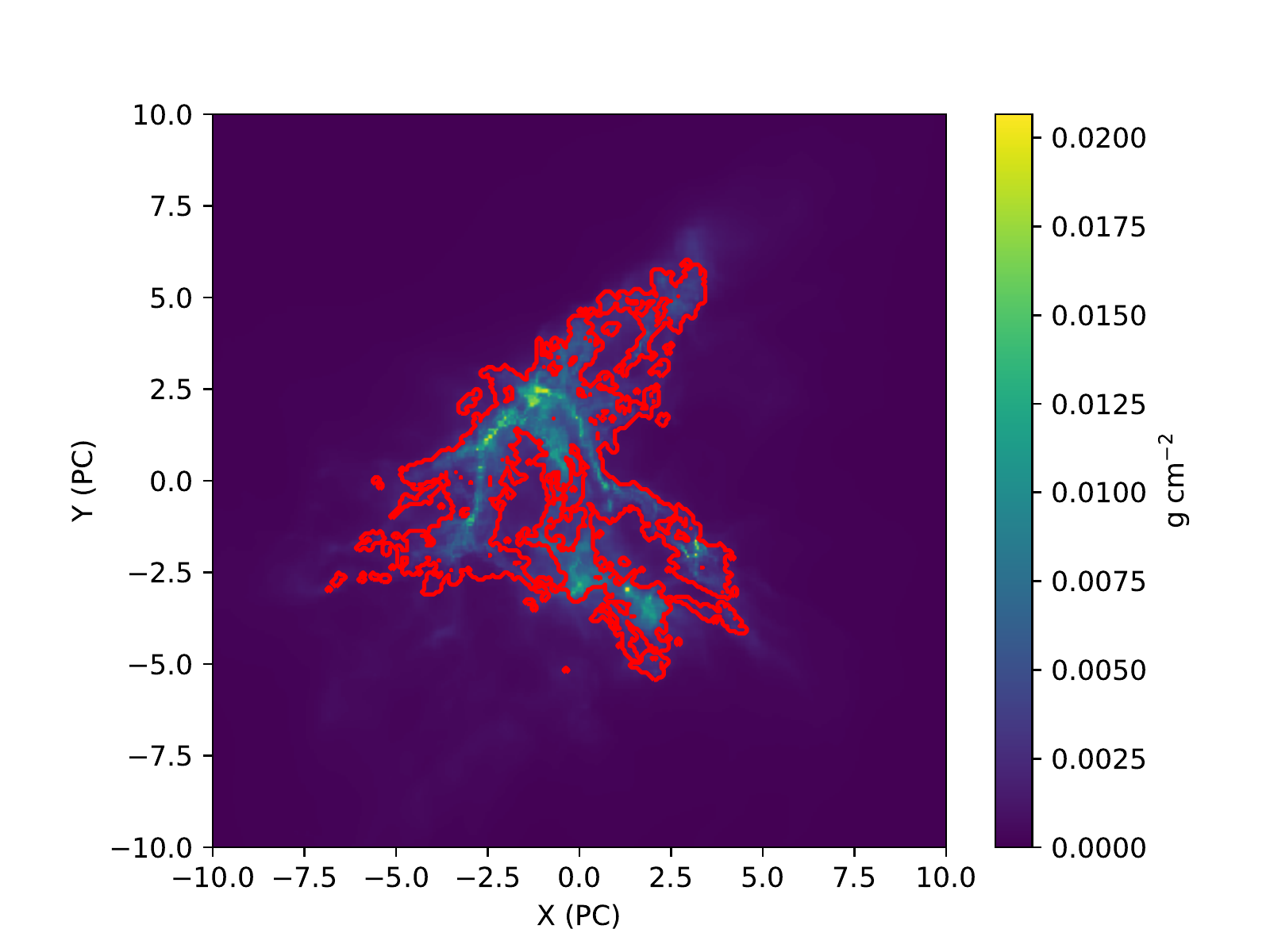}
  \includegraphics[width = 0.5 \textwidth]{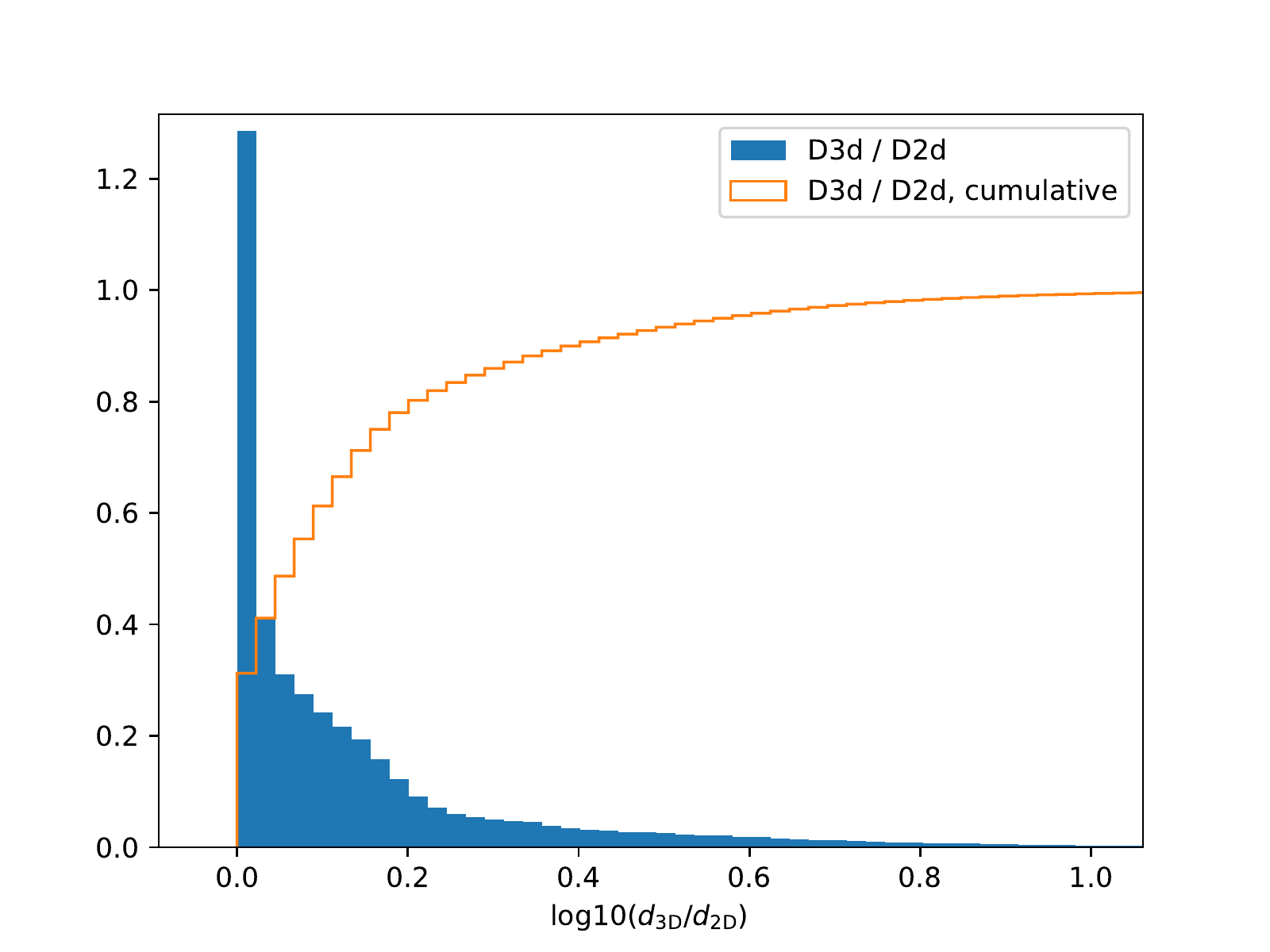}

  \caption{\label{fig:simu}Relation between physical distance and projection distance. The left panel plots the surface density distribution of the simulation from \citet{2019MNRAS.486.4622C}. The red contour outlines the region which contains gas with densities larger than  $n_{\rm H_2} \gtrsim 250\;\rm cm^{-3}$. This is the gas considered in our pairwise analysis. Right panel: distribution of the ratio between 3D physical distance and the projected distance for pairs of pixels. The blue bars represent the differential distribution and the yellow curve represents the cumulative distribution. The y-axis is normalized with respect to the cumulative distribution. 
  }
\end{figure*}

\citet{2019MNRAS.486.4622C} performed simulations addressing the formation of  a molecular cloud using the \texttt{Arepo} code 
\citep{2010MNRAS.401..791S}, where they simulated the formation of a 1300 $M_{\odot}$ cloud with self-consistent hydrogen, carbon, and oxygen chemistry. The surface density distribution of a simulated cloud is plotted in Fig. \ref{fig:simu}. Taking advantage of the simulation, we address the connection between 3D distance and 2D projected distance.
 
We start by choosing regions with gas densities above $n_{\rm H_2} \gtrsim 250\;\rm cm^{-3}$. The region within which the high-density gas lies is outlined in Fig. \ref{fig:simu}. We then plot the distribution of the ratio between the 3D physical distance and 2D projected distance $d_{\rm 2D} / d_{\rm 3D}$ for all pixel pairs, and estimated that 80  \% of the pairs has  {$d_{\rm 3D} / d_{\rm 2D} \lesssim 1.58$} . Thus, the projected distance should be a reasonable approximation to the physical distance.

\section{Predictions for turbulence and thermal-dominated fragmentation}\label{sec:appen:ana}
We use the techniques of dimensional analysis to derive the relation between core separation and surface density. In the case of thermal support, we assume that the core separation $l_{\rm core}$ is determined by the sound speed $c_{\rm s}$, the gravitational constant $G$ as well as the surface density $\Sigma_{\rm gas}$. By applying the Buckingham $\pi$ theorem, the governing equation of the system should take the form of
\begin{equation}
  f(c_{\rm s}^2 / (l_{\rm core} G \Sigma_{\rm gas})) = 0 \;,
\end{equation}
from which we expect 
\begin{equation}
  l_{\rm core, thermal} =  C \frac{c_{\rm s}^2}{G \Sigma_{\rm gas}} \approx \frac{c_{\rm s}^2}{G \Sigma_{\rm gas}} \,,
 \end{equation}
 where $C$ is a dimensionless number of order unity. This is the expected scaling between $l_{\rm core}$ and $\Sigma_{\rm gas}$ when the thermal support is effective.

 In the cases where a scale-dependent turbulence pressure support is effective, the turbulence can be parameterized in a model where the velocity dispersions $\sigma_{\rm v, turb}$ is related to scale $l$ by 
 \begin{equation}
  \sigma_{\rm v, turb} =  \sigma_0 (l / l_0)^{\beta}\;.
 \end{equation}
In this case, one can use the the dimensional number $\epsilon = \sigma_0^{1/\beta}  l_0^{-1}$ to describe the turbulence. In the case where $\beta = 1/3$, $\epsilon$ corresponds to the energy dissipation rate of the turbulence. The governing parameters of the system this consist of $l_{\rm core}$, $G$, $\Sigma_{\rm gas}$ and $\epsilon$, and the governing equation should be 
\begin{equation}
  f'(\Pi) =0, \, \Pi = \epsilon^{-2 \beta} G \Sigma_{\rm gas} l_{\rm core}^{1 - 2 \beta}\;.
\end{equation}
 Thus
 \begin{equation}
l_{\rm core, turbulent} \approx \epsilon^{\frac{-2 \beta}{2 \beta - 1}} G^{\frac{1}{2 \beta - 1}} \Sigma_{\rm gas} ^{\frac{1}{2 \beta -1}}\;.
 \end{equation}

\section{A colorblindness-proof version of Fig. 1}
\begin{figure*}[htb!]
  \epsscale{1.10}\plotone{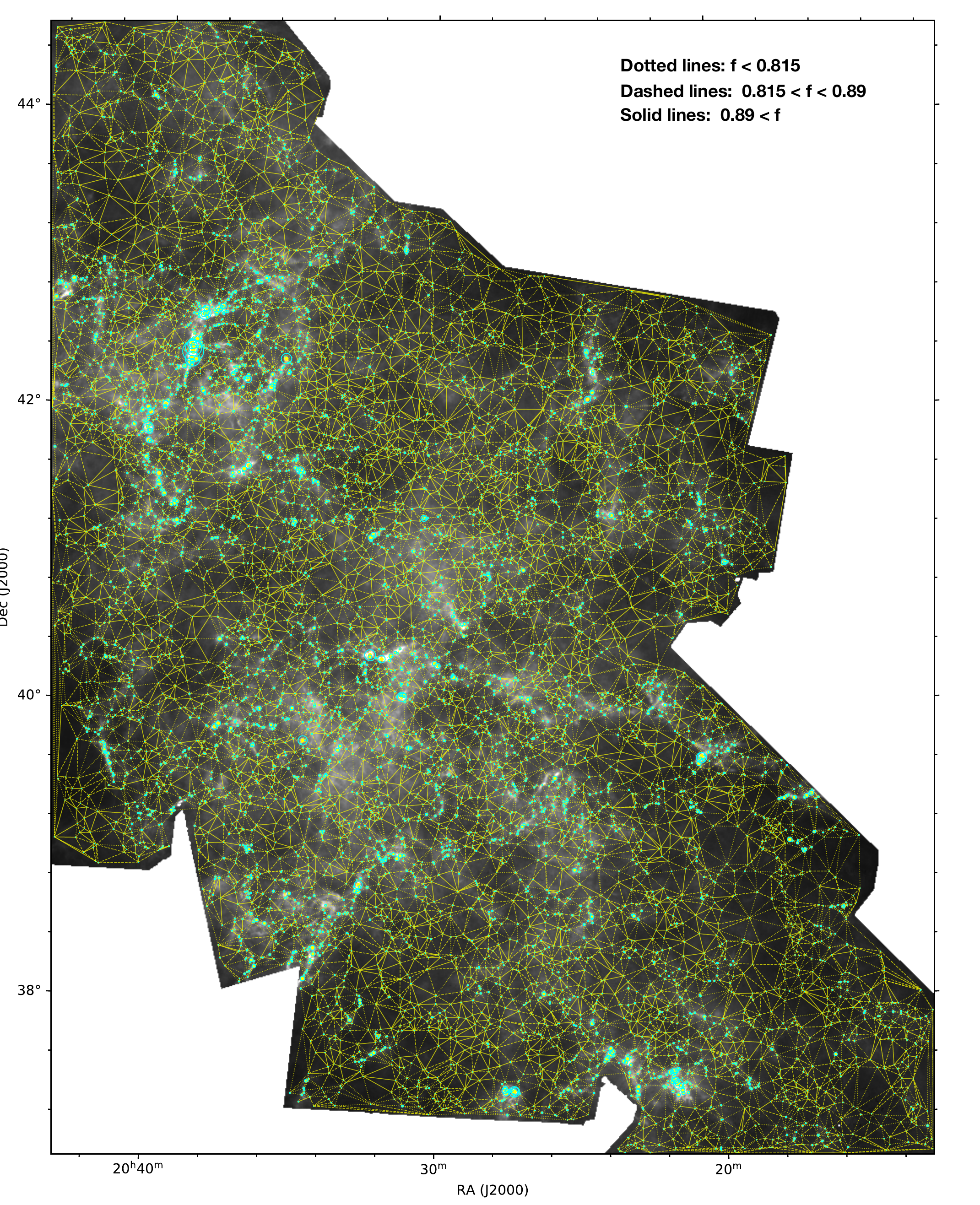}
   \caption{\label{fig:mesh2} {Delaunay triangulation constructed on the} 
   Cygnus-X GMC complex. The background image is a surface density map
   of the region constructed  using  data from  \citet{2010A&A...518L..77M}. On top of it, we overplot the dense core sample presented in Cao et al. (2021 {under review}), as well as the corresponding Delaunay
   triangulation diagram. {We use different line styles to represent edges of different filling factors} (Eq.(\ref{eq:filling})): the dotted lines stand for edges with filling factors smaller than 0.81, the dashed
   lines stand for edges with filling factors $0.815 < f < 0.89$, and the solid  lines stands for filling factors larger than 0.89. See Sec.
  \ref{sec:filling} for details. We also overlaid the accretion radii estimated using Eqs. (\ref{eq:acc:1}) and (\ref{eq:acc:2}). }
  \end{figure*}

Fig  \ref{fig:mesh2} we is a  colorblindness-proof version of Fig. \ref{fig:mesh}.

\acknowledgments

\facilities{Herschel}
\software{Scipy \citep{2020SciPy-NMeth}, Astropy \citep{2013A&A...558A..33A,2018AJ....156..123A}, Matplotlib \citep{Hunter:2007}, getsources \citep{Men2012}, Arepo \citep{2010MNRAS.401..791S}}

\bibliography{ms}{}

\begin{thebibliography}{}
\expandafter\ifx\csname natexlab\endcsname\relax\def\natexlab#1{#1}\fi
\providecommand{\url}[1]{\href{#1}{#1}}
\providecommand{\dodoi}[1]{doi:~\href{http://doi.org/#1}{\nolinkurl{#1}}}
\providecommand{\doeprint}[1]{\href{http://ascl.net/#1}{\nolinkurl{http://ascl.net/#1}}}
\providecommand{\doarXiv}[1]{\href{https://arxiv.org/abs/#1}{\nolinkurl{https://arxiv.org/abs/#1}}}

\bibitem[{{Alves} {et~al.}(2007){Alves}, {Lombardi}, \&
  {Lada}}]{2007A&A...462L..17A}
{Alves}, J., {Lombardi}, M., \& {Lada}, C.~J. 2007, \aap, 462, L17,
  \dodoi{10.1051/0004-6361:20066389}

\bibitem[{{Alves} {et~al.}(2001){Alves}, {Lada}, \&
  {Lada}}]{2001Natur.409..159A}
{Alves}, J.~F., {Lada}, C.~J., \& {Lada}, E.~A. 2001, \nat, 409, 159

\bibitem[{{Andr{\'e}} {et~al.}(2014){Andr{\'e}}, {Di Francesco},
  {Ward-Thompson}, {Inutsuka}, {Pudritz}, \& {Pineda}}]{2014prpl.conf...27A}
{Andr{\'e}}, P., {Di Francesco}, J., {Ward-Thompson}, D., {et~al.} 2014, {From
  Filamentary Networks to Dense Cores in Molecular Clouds: Toward a New
  Paradigm for Star Formation} (Protostars and Planets VI, Tucson, AZ, Univ.
  Arizona Press)), 27--51, \dodoi{10.2458/azu_uapress_9780816531240-ch002}

\bibitem[{{Astropy Collaboration} {et~al.}(2013){Astropy Collaboration},
  {Robitaille}, {Tollerud}, {Greenfield}, {Droettboom}, {Bray}, {Aldcroft},
  {Davis}, {Ginsburg}, {Price-Whelan}, {Kerzendorf}, {Conley}, {Crighton},
  {Barbary}, {Muna}, {Ferguson}, {Grollier}, {Parikh}, {Nair}, {Unther},
  {Deil}, {Woillez}, {Conseil}, {Kramer}, {Turner}, {Singer}, {Fox}, {Weaver},
  {Zabalza}, {Edwards}, {Azalee Bostroem}, {Burke}, {Casey}, {Crawford},
  {Dencheva}, {Ely}, {Jenness}, {Labrie}, {Lim}, {Pierfederici}, {Pontzen},
  {Ptak}, {Refsdal}, {Servillat}, \& {Streicher}}]{2013A&A...558A..33A}
{Astropy Collaboration}, {Robitaille}, T.~P., {Tollerud}, E.~J., {et~al.} 2013,
  \aap, 558, A33, \dodoi{10.1051/0004-6361/201322068}

\bibitem[{{Astropy Collaboration} {et~al.}(2018){Astropy Collaboration},
  {Price-Whelan}, {Sip{\H{o}}cz}, {G{\"u}nther}, {Lim}, {Crawford}, {Conseil},
  {Shupe}, {Craig}, {Dencheva}, {Ginsburg}, {VanderPlas}, {Bradley},
  {P{\'e}rez-Su{\'a}rez}, {de Val-Borro}, {Aldcroft}, {Cruz}, {Robitaille},
  {Tollerud}, {Ardelean}, {Babej}, {Bach}, {Bachetti}, {Bakanov}, {Bamford},
  {Barentsen}, {Barmby}, {Baumbach}, {Berry}, {Biscani}, {Boquien}, {Bostroem},
  {Bouma}, {Brammer}, {Bray}, {Breytenbach}, {Buddelmeijer}, {Burke},
  {Calderone}, {Cano Rodr{\'\i}guez}, {Cara}, {Cardoso}, {Cheedella}, {Copin},
  {Corrales}, {Crichton}, {D'Avella}, {Deil}, {Depagne}, {Dietrich}, {Donath},
  {Droettboom}, {Earl}, {Erben}, {Fabbro}, {Ferreira}, {Finethy}, {Fox},
  {Garrison}, {Gibbons}, {Goldstein}, {Gommers}, {Greco}, {Greenfield},
  {Groener}, {Grollier}, {Hagen}, {Hirst}, {Homeier}, {Horton}, {Hosseinzadeh},
  {Hu}, {Hunkeler}, {Ivezi{\'c}}, {Jain}, {Jenness}, {Kanarek}, {Kendrew},
  {Kern}, {Kerzendorf}, {Khvalko}, {King}, {Kirkby}, {Kulkarni}, {Kumar},
  {Lee}, {Lenz}, {Littlefair}, {Ma}, {Macleod}, {Mastropietro}, {McCully},
  {Montagnac}, {Morris}, {Mueller}, {Mumford}, {Muna}, {Murphy}, {Nelson},
  {Nguyen}, {Ninan}, {N{\"o}the}, {Ogaz}, {Oh}, {Parejko}, {Parley}, {Pascual},
  {Patil}, {Patil}, {Plunkett}, {Prochaska}, {Rastogi}, {Reddy Janga},
  {Sabater}, {Sakurikar}, {Seifert}, {Sherbert}, {Sherwood-Taylor}, {Shih},
  {Sick}, {Silbiger}, {Singanamalla}, {Singer}, {Sladen}, {Sooley},
  {Sornarajah}, {Streicher}, {Teuben}, {Thomas}, {Tremblay}, {Turner},
  {Terr{\'o}n}, {van Kerkwijk}, {de la Vega}, {Watkins}, {Weaver}, {Whitmore},
  {Woillez}, {Zabalza}, \& {Astropy Contributors}}]{2018AJ....156..123A}
{Astropy Collaboration}, {Price-Whelan}, A.~M., {Sip{\H{o}}cz}, B.~M., {et~al.}
  2018, \aj, 156, 123, \dodoi{10.3847/1538-3881/aabc4f}

\bibitem[{{Ballesteros-Paredes} {et~al.}(2011){Ballesteros-Paredes},
  {Hartmann}, {V{\'a}zquez-Semadeni}, {Heitsch}, \&
  {Zamora-Avil{\'e}s}}]{2011MNRAS.411...65B}
{Ballesteros-Paredes}, J., {Hartmann}, L.~W., {V{\'a}zquez-Semadeni}, E.,
  {Heitsch}, F., \& {Zamora-Avil{\'e}s}, M.~A. 2011, \mnras, 411, 65,
  \dodoi{10.1111/j.1365-2966.2010.17657.x}

\bibitem[{{Ballesteros-Paredes} {et~al.}(2019){Ballesteros-Paredes},
  {Rom{\'a}n-Z{\'u}{\~n}iga}, {Salom{\'e}}, {Zamora-Avil{\'e}s}, \&
  {Jim{\'e}nez-Donaire}}]{2019MNRAS.490.2648B}
{Ballesteros-Paredes}, J., {Rom{\'a}n-Z{\'u}{\~n}iga}, C., {Salom{\'e}}, Q.,
  {Zamora-Avil{\'e}s}, M., \& {Jim{\'e}nez-Donaire}, M.~J. 2019, \mnras, 490,
  2648, \dodoi{10.1093/mnras/stz2575}

\bibitem[{{Beuther} {et~al.}(2007){Beuther}, {Churchwell}, {McKee}, \&
  {Tan}}]{2007prpl.conf..165B}
{Beuther}, H., {Churchwell}, E.~B., {McKee}, C.~F., \& {Tan}, J.~C. 2007, {The
  Formation of Massive Stars}, ed. B.~{Reipurth}, D.~{Jewitt}, \& K.~{Keil}
  (Protostars and Planets V, Tucson, AZ, Univ. Arizona Press), 165.
\newblock \doarXiv{astro-ph/0602012}

\bibitem[{{Bonnell} \& {Bate}(2006)}]{2006MNRAS.370..488B}
{Bonnell}, I.~A., \& {Bate}, M.~R. 2006, \mnras, 370, 488,
  \dodoi{10.1111/j.1365-2966.2006.10495.x}

\bibitem[{{Burkhart} {et~al.}(2017){Burkhart}, {Stalpes}, \&
  {Collins}}]{2017ApJ...834L...1B}
{Burkhart}, B., {Stalpes}, K., \& {Collins}, D.~C. 2017, \apjl, 834, L1,
  \dodoi{10.3847/2041-8213/834/1/L1}

\bibitem[{{Cao} {et~al.}(2019){Cao}, {Qiu}, {Zhang}, {Wang}, {Hu}, \&
  {Liu}}]{2019ApJS..241....1C}
{Cao}, Y., {Qiu}, K., {Zhang}, Q., {et~al.} 2019, \apjs, 241, 1,
  \dodoi{10.3847/1538-4365/ab0025}

\bibitem[{{Chira} {et~al.}(2019){Chira}, {Ib{\'a}{\~n}ez-Mej{\'\i}a}, {Mac
  Low}, \& {Henning}}]{2019A&A...630A..97C}
{Chira}, R.~A., {Ib{\'a}{\~n}ez-Mej{\'\i}a}, J.~C., {Mac Low}, M.~M., \&
  {Henning}, T. 2019, \aap, 630, A97, \dodoi{10.1051/0004-6361/201833970}

\bibitem[{{Clark} {et~al.}(2019){Clark}, {Glover}, {Ragan}, \&
  {Duarte-Cabral}}]{2019MNRAS.486.4622C}
{Clark}, P.~C., {Glover}, S. C.~O., {Ragan}, S.~E., \& {Duarte-Cabral}, A.
  2019, \mnras, 486, 4622, \dodoi{10.1093/mnras/stz1119}

\bibitem[{{Dale} \& {Bonnell}(2011)}]{2011MNRAS.414..321D}
{Dale}, J.~E., \& {Bonnell}, I. 2011, \mnras, 414, 321,
  \dodoi{10.1111/j.1365-2966.2011.18392.x}

\bibitem[{Delaunay(1934)}]{delaunay1934}
Delaunay, B. 1934, Izv. Akad. Nauk SSSR, Otdelenie Matematicheskii i
  Estestvennyka Nauk, 7, 793

\bibitem[{{Dobbs} {et~al.}(2014){Dobbs}, {Krumholz}, {Ballesteros-Paredes},
  {Bolatto}, {Fukui}, {Heyer}, {Low}, {Ostriker}, \&
  {V{\'a}zquez-Semadeni}}]{2014prpl.conf....3D}
{Dobbs}, C.~L., {Krumholz}, M.~R., {Ballesteros-Paredes}, J., {et~al.} 2014,
  {Formation of Molecular Clouds and Global Conditions for Star Formation}, ed.
  H.~{Beuther}, R.~S. {Klessen}, C.~P. {Dullemond}, \& T.~{Henning} (Protostars
  and Planets VI, Tucson, AZ, Univ. Arizona Press)), 3,
  \dodoi{10.2458/azu_uapress_9780816531240-ch001}

\bibitem[{{Goodwin} \& {Kroupa}(2005)}]{2005A&A...439..565G}
{Goodwin}, S.~P., \& {Kroupa}, P. 2005, \aap, 439, 565,
  \dodoi{10.1051/0004-6361:20052654}

\bibitem[{{Guerrero-Gamboa} \&
  {V{\'a}zquez-Semadeni}(2020)}]{2020arXiv200904676G}
{Guerrero-Gamboa}, R., \& {V{\'a}zquez-Semadeni}, E. 2020, arXiv e-prints,
  arXiv:2009.04676.
\newblock \doarXiv{2009.04676}

\bibitem[{{Hennebelle} \& {Chabrier}(2009)}]{2009ApJ...702.1428H}
{Hennebelle}, P., \& {Chabrier}, G. 2009, \apj, 702, 1428,
  \dodoi{10.1088/0004-637X/702/2/1428}

\bibitem[{{Heyer} {et~al.}(2009){Heyer}, {Krawczyk}, {Duval}, \&
  {Jackson}}]{2009ApJ...699.1092H}
{Heyer}, M., {Krawczyk}, C., {Duval}, J., \& {Jackson}, J.~M. 2009, \apj, 699,
  1092, \dodoi{10.1088/0004-637X/699/2/1092}

\bibitem[{{Heyer} \& {Brunt}(2004)}]{2004ApJ...615L..45H}
{Heyer}, M.~H., \& {Brunt}, C.~M. 2004, \apjl, 615, L45, \dodoi{10.1086/425978}

\bibitem[{{Hopkins}(2012)}]{2012MNRAS.423.2037H}
{Hopkins}, P.~F. 2012, \mnras, 423, 2037,
  \dodoi{10.1111/j.1365-2966.2012.20731.x}

\bibitem[{Hunter(2007)}]{Hunter:2007}
Hunter, J.~D. 2007, Computing in Science \& Engineering, 9, 90,
  \dodoi{10.1109/MCSE.2007.55}

\bibitem[{{Kainulainen} {et~al.}(2009){Kainulainen}, {Beuther}, {Henning}, \&
  {Plume}}]{2009A&A...508L..35K}
{Kainulainen}, J., {Beuther}, H., {Henning}, T., \& {Plume}, R. 2009, \aap,
  508, L35, \dodoi{10.1051/0004-6361/200913605}

\bibitem[{{Klessen}(2000)}]{2000ApJ...535..869K}
{Klessen}, R.~S. 2000, \apj, 535, 869, \dodoi{10.1086/308854}

\bibitem[{{Kritsuk} {et~al.}(2011){Kritsuk}, {Norman}, \&
  {Wagner}}]{2011ApJ...727L..20K}
{Kritsuk}, A.~G., {Norman}, M.~L., \& {Wagner}, R. 2011, \apjl, 727, L20,
  \dodoi{10.1088/2041-8205/727/1/L20}

\bibitem[{{Larson}(1981)}]{1981MNRAS.194..809L}
{Larson}, R.~B. 1981, \mnras, 194, 809, \dodoi{10.1093/mnras/194.4.809}

\bibitem[{{Li}(2017)}]{Li2017}
{Li}, G.-X. 2017, \mnras, 465, 667, \dodoi{10.1093/mnras/stw2707}

\bibitem[{{Li}(2018)}]{2018MNRAS.477.4951L}
---. 2018, \mnras, 477, 4951, \dodoi{10.1093/mnras/sty657}

\bibitem[{{Li} {et~al.}(2016{\natexlab{a}}){Li}, {Burkert}, {Megeath}, \&
  {Wyrowski}}]{2016arXiv160305720L}
{Li}, G.-X., {Burkert}, A., {Megeath}, T., \& {Wyrowski}, F.
  2016{\natexlab{a}}, arXiv e-prints, arXiv:1603.05720.
\newblock \doarXiv{1603.05720}

\bibitem[{{Li} {et~al.}(2016{\natexlab{b}}){Li}, {Urquhart}, {Leurini},
  {Csengeri}, {Wyrowski}, {Menten}, \& {Schuller}}]{2016A&A...591A...5L}
{Li}, G.-X., {Urquhart}, J.~S., {Leurini}, S., {et~al.} 2016{\natexlab{b}},
  \aap, 591, A5, \dodoi{10.1051/0004-6361/201527468}

\bibitem[{{Li} {et~al.}(2014){Li}, {Goodman}, {Sridharan}, {Houde}, {Li},
  {Novak}, \& {Tang}}]{2014prpl.conf..101L}
{Li}, H.-B., {Goodman}, A., {Sridharan}, T.~K., {et~al.} 2014, {The Link
  Between Magnetic Fields and Cloud/Star Formation} (Protostars and Planets VI,
  Tucson, AZ, Univ. Arizona Press), 101--123,
  \dodoi{10.2458/azu_uapress_9780816531240-ch005}

\bibitem[{{Lu} {et~al.}(2018){Lu}, {Zhang}, {Liu}, {Sanhueza}, {Tatematsu},
  {Feng}, {Smith}, {Myers}, {Sridharan}, \& {Gu}}]{2018ApJ...855....9L}
{Lu}, X., {Zhang}, Q., {Liu}, H.~B., {et~al.} 2018, \apj, 855, 9,
  \dodoi{10.3847/1538-4357/aaad11}

\bibitem[{{Mac Low} \& {Klessen}(2004)}]{2004RvMP...76..125M}
{Mac Low}, M.-M., \& {Klessen}, R.~S. 2004, Reviews of Modern Physics, 76, 125,
  \dodoi{10.1103/RevModPhys.76.125}

\bibitem[{{McKee} \& {Tan}(2003)}]{2003ApJ...585..850M}
{McKee}, C.~F., \& {Tan}, J.~C. 2003, \apj, 585, 850, \dodoi{10.1086/346149}

\bibitem[{{Men'shchikov} {et~al.}(2012{\natexlab{a}}){Men'shchikov},
  {Andr{\'e}}, {Didelon}, {Motte}, {Hennemann}, \&
  {Schneider}}]{2012A&A...542A..81M}
{Men'shchikov}, A., {Andr{\'e}}, P., {Didelon}, P., {et~al.}
  2012{\natexlab{a}}, \aap, 542, A81, \dodoi{10.1051/0004-6361/201218797}

\bibitem[{{Men'shchikov} {et~al.}(2012{\natexlab{b}}){Men'shchikov},
  {Andr{\'e}}, {Didelon}, {Motte}, {Hennemann}, \& {Schneider}}]{Men2012}
---. 2012{\natexlab{b}}, \aap, 542, A81, \dodoi{10.1051/0004-6361/201218797}

\bibitem[{{Motte} {et~al.}(2007){Motte}, {Bontemps}, {Schilke}, {Schneider},
  {Menten}, \& {Brogui{\`e}re}}]{2007A&A...476.1243M}
{Motte}, F., {Bontemps}, S., {Schilke}, P., {et~al.} 2007, \aap, 476, 1243,
  \dodoi{10.1051/0004-6361:20077843}

\bibitem[{{Motte} {et~al.}(2010){Motte}, {Zavagno}, {Bontemps}, {Schneider},
  {Hennemann}, {di Francesco}, {Andr{\'e}}, {Saraceno}, {Griffin}, {Marston},
  {Ward-Thompson}, {White}, {Minier}, {Men'shchikov}, {Hill}, {Abergel},
  {Anderson}, {Aussel}, {Balog}, {Baluteau}, {Bernard}, {Cox}, {Csengeri},
  {Deharveng}, {Didelon}, {di Giorgio}, {Hargrave}, {Huang}, {Kirk}, {Leeks},
  {Li}, {Martin}, {Molinari}, {Nguyen-Luong}, {Olofsson}, {Persi}, {Peretto},
  {Pezzuto}, {Roussel}, {Russeil}, {Sadavoy}, {Sauvage}, {Sibthorpe},
  {Spinoglio}, {Testi}, {Teyssier}, {Vavrek}, {Wilson}, \&
  {Woodcraft}}]{2010A&A...518L..77M}
{Motte}, F., {Zavagno}, A., {Bontemps}, S., {et~al.} 2010, \aap, 518, L77,
  \dodoi{10.1051/0004-6361/201014690}

\bibitem[{{Murray} \& {Chang}(2012)}]{2012ApJ...746...75M}
{Murray}, N., \& {Chang}, P. 2012, \apj, 746, 75,
  \dodoi{10.1088/0004-637X/746/1/75}

\bibitem[{{Murray} \& {Chang}(2015)}]{2015ApJ...804...44M}
---. 2015, \apj, 804, 44, \dodoi{10.1088/0004-637X/804/1/44}

\bibitem[{{Padoan} \& {Nordlund}(2002)}]{2002ApJ...576..870P}
{Padoan}, P., \& {Nordlund}, {\r{A}}. 2002, \apj, 576, 870,
  \dodoi{10.1086/341790}

\bibitem[{{Palau} {et~al.}(2015){Palau}, {Ballesteros-Paredes},
  {V{\'a}zquez-Semadeni}, {S{\'a}nchez-Monge}, {Estalella}, {Fall}, {Zapata},
  {Camacho}, {G{\'o}mez}, {Naranjo-Romero}, {Busquet}, \&
  {Fontani}}]{2015MNRAS.453.3785P}
{Palau}, A., {Ballesteros-Paredes}, J., {V{\'a}zquez-Semadeni}, E., {et~al.}
  2015, \mnras, 453, 3785, \dodoi{10.1093/mnras/stv1834}

\bibitem[{{Palau} {et~al.}(2018){Palau}, {Zapata}, {Rom{\'a}n-Z{\'u}{\~n}iga},
  {S{\'a}nchez-Monge}, {Estalella}, {Busquet}, {Girart}, {Fuente}, \&
  {Commer{\c{c}}on}}]{2018ApJ...855...24P}
{Palau}, A., {Zapata}, L.~A., {Rom{\'a}n-Z{\'u}{\~n}iga}, C.~G., {et~al.} 2018,
  \apj, 855, 24, \dodoi{10.3847/1538-4357/aaad03}

\bibitem[{{Qian} {et~al.}(2018){Qian}, {Li}, {Gao}, {Xu}, \&
  {Pan}}]{2018ApJ...864..116Q}
{Qian}, L., {Li}, D., {Gao}, Y., {Xu}, H., \& {Pan}, Z. 2018, \apj, 864, 116,
  \dodoi{10.3847/1538-4357/aad780}

\bibitem[{{Robertson} \& {Goldreich}(2012)}]{2012ApJ...750L..31R}
{Robertson}, B., \& {Goldreich}, P. 2012, \apjl, 750, L31,
  \dodoi{10.1088/2041-8205/750/2/L31}

\bibitem[{{Roman-Duval} {et~al.}(2011){Roman-Duval}, {Federrath}, {Brunt},
  {Heyer}, {Jackson}, \& {Klessen}}]{2011ApJ...740..120R}
{Roman-Duval}, J., {Federrath}, C., {Brunt}, C., {et~al.} 2011, \apj, 740, 120,
  \dodoi{10.1088/0004-637X/740/2/120}

\bibitem[{{Rygl} {et~al.}(2012){Rygl}, {Brunthaler}, {Sanna}, {Menten}, {Reid},
  {van Langevelde}, {Honma}, {Torstensson}, \&
  {Fujisawa}}]{2012A&A...539A..79R}
{Rygl}, K.~L.~J., {Brunthaler}, A., {Sanna}, A., {et~al.} 2012, \aap, 539, A79,
  \dodoi{10.1051/0004-6361/201118211}

\bibitem[{{Springel}(2010)}]{2010MNRAS.401..791S}
{Springel}, V. 2010, \mnras, 401, 791, \dodoi{10.1111/j.1365-2966.2009.15715.x}

\bibitem[{{Tig{\'e}} {et~al.}(2017){Tig{\'e}}, {Motte}, {Russeil}, {Zavagno},
  {Hennemann}, {Schneider}, {Hill}, {Nguyen Luong}, {Di Francesco}, {Bontemps},
  {Louvet}, {Didelon}, {K{\"o}nyves}, {Andr{\'e}}, {Leuleu}, {Bardagi},
  {Anderson}, {Arzoumanian}, {Benedettini}, {Bernard}, {Elia}, {Figueira},
  {Kirk}, {Martin}, {Minier}, {Molinari}, {Nony}, {Persi}, {Pezzuto},
  {Polychroni}, {Rayner}, {Rivera-Ingraham}, {Roussel}, {Rygl}, {Spinoglio}, \&
  {White}}]{2017A&A...602A..77T}
{Tig{\'e}}, J., {Motte}, F., {Russeil}, D., {et~al.} 2017, \aap, 602, A77,
  \dodoi{10.1051/0004-6361/201628989}

\bibitem[{Virtanen {et~al.}(2020)Virtanen, Gommers, Oliphant, Haberland, Reddy,
  Cournapeau, Burovski, Peterson, Weckesser, Bright, {van der Walt}, Brett,
  Wilson, Millman, Mayorov, Nelson, Jones, Kern, Larson, Carey, Polat, Feng,
  Moore, {VanderPlas}, Laxalde, Perktold, Cimrman, Henriksen, Quintero, Harris,
  Archibald, Ribeiro, Pedregosa, {van Mulbregt}, \& {SciPy 1.0
  Contributors}}]{2020SciPy-NMeth}
Virtanen, P., Gommers, R., Oliphant, T.~E., {et~al.} 2020, Nature Methods, 17,
  261, \dodoi{10.1038/s41592-019-0686-2}

\bibitem[{{Williams} {et~al.}(2000){Williams}, {Blitz}, \&
  {McKee}}]{Williams2000}
{Williams}, J.~P., {Blitz}, L., \& {McKee}, C.~F. 2000, {The Structure and
  Evolution of Molecular Clouds: from Clumps to Cores to the IMF} (Protostars
  and Planets IV, Tucson, AZ, Univ. Arizona Press)), 97

\bibitem[{{Zhang} \& {Li}(2017)}]{Zhang2017b}
{Zhang}, C.-P., \& {Li}, G.-X. 2017, \mnras, 469, 2286,
  \dodoi{10.1093/mnras/stx954}

\bibitem[{{Zhang} {et~al.}(2009){Zhang}, {Wang}, {Pillai}, \&
  {Rathborne}}]{2009ApJ...696..268Z}
{Zhang}, Q., {Wang}, Y., {Pillai}, T., \& {Rathborne}, J. 2009, \apj, 696, 268,
  \dodoi{10.1088/0004-637X/696/1/268}

\end{thebibliography}
\bibliographystyle{aasjournal}


\end{document}